\documentclass{multi3}  
\begin{document}
\newcommand{\half}{\mbox{$\textstyle \frac{1}{2}$}}
\newcommand{\ket}[1]{\left | \, #1 \right \rangle}
\newcommand{\bra}[1]{\left \langle #1 \, \right |}
\newcommand{\proj}[1]{\ket{#1}\bra{#1}}
\newcommand{\av}[1]{\langle #1\rangle}
\newcommand{\beq}{\begin{equation}}
\newcommand{\eeq}{\end{equation}}

\input{psfig.sty}

\chapter{From quantum code-making to quantum code-breaking}
\author{Artur Ekert}
\address{Clarendon Laboratory and Merton College, University of Oxford.}

\section{What is wrong with classical cryptography ?}

Human desire to communicate secretly is at least as old as writing itself and goes back
to the beginnings of our civilisation. Methods of secret communication were developed
by many ancient societies, including those of Mesopotamia, Egypt, India, and China, but
details regarding the origins of cryptology\footnote{The science of secure
communication is called cryptology from Greek {\em kryptos} hidden and {\em logos}
word. Cryptology embodies cryptography, the art of code-making, and cryptanalysis, the
art of code-breaking.} remain unknown (Kahn 1967).

We know that it was the Spartans, the most warlike of the Greeks, who pioneered
military cryptography in Europe. Around 400 BC they employed a device known as a {\it
the scytale}. The device, used for communication between military commanders,
consisted of a tapered baton around which was wrapped a spiral  strip of parchment or
leather containing the message. Words were then written lengthwise along the baton,
one letter on each revolution of the strip. When unwrapped, the letters of the message
appeared scrambled and the parchment was sent on its way. The receiver wrapped the
parchment around another baton of the same shape and the original message reappeared.

Julius Caesar allegedly used, in his correspondence, a simple letter substitution method.
Each letter of Caesar's message was replaced by the letter that followed it
alphabetically by three places. The letter A was replaced by D, the letter B by E, and so
on. For example, the English word COLD after the Caesar substitution appears as FROG.
This method is still called the Caesar cipher, regardless the size of the shift used for
the substitution.

These two simple examples already contain the two basic methods of encryption which
are still employed by cryptographers today namely {\em transposition} and {\em
substitution}. In transposition ({\em e.g. scytale}) the letters of the {\em plaintext}, the
technical term for the message to be transmitted, are rearranged by a special
permutation. In substitution ({\em e.g.} Caesar's cipher) the letters of the plaintext are
replaced by other letters, numbers or arbitrary symbols. In general the two techniques
can be combined (for an introduction to modern cryptology see, for example,  (Menzes et
al. 1996; Schneier 1994; Welsh 1988).

Originally the security of a cryptotext depended on the secrecy of the entire encrypting
and decrypting procedures; however, today we use ciphers for which the algorithm for
encrypting and decrypting could be revealed to anybody without compromising the
security of a particular cryptogram. In such ciphers a set of specific parameters, called
a {\it key}, is supplied together with the plaintext as an input to the encrypting
algorithm, and together with the cryptogram as an input to the decrypting algorithm.
This can be written as
\begin{equation}
\hat E_k(P) = C,  \; {\rm and\; conversely,}\; \hat D_k(C)=P,
\end{equation}
where $P$ stands for plaintext, $C$ for cryptotext or cryptogram, $k$ for cryptographic
key, and $\hat E$ and $\hat D$ denote an encryption and a decryption operation
respectively.

The encrypting and decrypting algorithms are publicly known; the security of the
cryptogram depends entirely on the secrecy of the key, and this key must consist of
a {\it randomly chosen}, sufficiently long string of bits. Probably the best way to
explain this procedure is to have a quick look at the Vernam cipher, also known as the
one-time pad.

If we choose a very simple digital alphabet in which we use only capital letters and
some punctuation marks such as

\bigskip
\begin{center}
\begin{tabular}{|ccccccccccccccc|}
\hline 
A & B & C & D & E & ... & \  & ... & X & Y & Z & \  & ? & , & . \\
01 & 02 & 03 & 04 & 05 & ... & \  & ... & 24 & 25 & 26 & 27 & 28 & 29 &
30 \\
\hline
\end{tabular}
\end{center}
\bigskip

we can illustrate the secret-key encrypting procedure by the following simple example:

\begin{center}
{\small
\begin{tabular}{|cccccccccccc|}
\hline
H&E&L&L&O&\ &R&O&G&E&R&. \\
08&05&12&12&15&27&18&15&07&05&18&30\\
24&14&26&25&29&17&28&12&01&18&27&03\\
02&19&08&07&14&14&16&07&08&23&15&03\\
\hline
\end{tabular}
}
\end{center}

In order to obtain the cryptogram (sequence of digits in the bottom row) we add the
plaintext numbers (the top row of digits) to the key numbers (the middle row), which are
randomly selected from between 1 and 30, and take the remainder after division of the
sum by 30, that is we perform addition modulo 30. For example, the first letter of the
message ``H" becomes a number \lq\lq 08\rq\rq in the plaintext, then we add $ 08 + 24 =
32 \ ; \ 32=1\times 30 + 2 $, therefore we get 02 in the cryptogram. The encryption and
decryption can be written as $P+k\; ({\rm mod}\; 30) = C$ and $C-k\; ({\rm mod}\; 30)
=P$ respectively. 

The cipher was invented by Major Joseph Mauborgne and AT\&T's Gilbert Vernam in 1917
and we know that if the key is secure, the same length as the message, truly random, and
never reused, this cipher is really unbreakable! So what is wrong with classical
cryptography?
 
There is a snag. It is called {\em key distribution}. Once the key is established,
subsequent communication involves sending cryptograms over a channel, even one which
is vulnerable to total passive eavesdropping (e.g. public announcement in mass-media).
However in order to establish the key, two users, who share no secret information
initially, must at a certain stage of communication use a reliable and a very secure
channel. Since the interception is a set of measurements performed by the eavesdropper
on this channel, however difficult this might be from a technological point of view, {\it
in principle} any {classical} key distribution can always be passively monitored, without
the legitimate users being aware that any eavesdropping has taken place. 

Cryptologists have tried hard to solve the key distribution problem. The 1970s, for
example, brought a clever mathematical discovery in the shape of ``public key" systems
(Diffie and Hellman 1977). In these systems users do not need to agree on a secret key
before they send the message. They work on the principle of a safe with two keys, one
public key to lock it, and another private one to open it. Everyone has a key to lock the
safe but only one person has a key that will open it again, so anyone can put a message in
the safe but only one person can take it out. These systems exploit the fact that certain
mathematical operations are easier to do in one direction than the other. The systems
avoid the key distribution problem but unfortunately their security depends on unproven
mathematical assumptions, such as the difficulty of factoring large
integers.\footnote{RSA - a very popular public key cryptosystem named after the
three inventors, Ron Rivest, Adi Shamir, and Leonard Adleman (1979) - gets its security
from the difficulty of factoring large numbers.} This means that if and when
mathematicians or computer scientists come up with fast and clever procedures for
factoring large integers the whole privacy and discretion of public-key cryptosystems
could vanish overnight. Indeed, recent work in quantum computation shows that quantum
computers can, at least in principle, factor much faster than classical computers (Shor
1994) !

In the following I will describe how quantum entanglement, singled out by Erwin
Schr\"odinger (1935) as the most remarkable feature of quantum theory, became an
important resource in the new field of quantum data processing. After a brief outline of
entanglement's key role in philosophical debates about the meaning of quantum
mechanics I will describe its current impact on both cryptography and cryptanalysis.
Thus this is a story about quantum code-making and quantum code-breaking.

\section{Is the Bell theorem of any practical use ?}

Probably the best way to agitate a group of jaded but philosophically-inclined physicists
is to buy them a bottle of wine and mention {\em interpretations of quantum mechanics}.
It is like opening Pandora's box. It seems that everybody agrees with the formalism of
quantum mechanics, but no one agrees on its meaning.  This is despite the fact that, as
far as lip-service goes, one particular orthodoxy established by Niels Bohr over 50 years
ago and known as the ``Copenhagen interpretation" still effectively holds sway.  It has
never been clear to me how so many physicists can seriously endorse a view according to
which the equations of quantum theory (e.g. the Schr\"odinger equation) apply only to
un-observed physical phenomena, while at the moment of observation a completely
different and mysterious process takes over.  Quantum theory, according to this view,
provides merely a calculational procedure and does not attempt to describe  {\em
objective physical reality}. A very defeatist view indeed.

One of the first who found the pragmatic instrumentalism of Bohr unacceptable was
Albert Einstein who, in 1927 during the fifth Solvay Conference in Brussels,  directly
challenged Bohr over the meaning of quantum theory. The intelectual atmosphere and the
philosophy of science at the time were dominated by positivism which gave Bohr the
edge, but Einstein stuck to his guns and the Bohr-Einstein debate lasted almost three
decades (after all they could not  use e-mail). In 1935 Einstein together with Boris
Podolsky and Nathan Rosen (EPR) published a paper in which they outlined how a `proper'
fundamental theory of nature should look like (Einstein et al. 1935).  The EPR programme
required completeness (``In a complete theory there is an element corresponding to each
element of reality"),  locality (``The real factual situation of the system A is independent
of what is done with the system B, which is spatially separated from the former"), and
defined the element of physical reality as ``If, without in any way disturbing a system,
we can predict with certainty the value of a physical quantity, then there exists an
element of physical reality corresponding to this physical quantity". EPR then considered
a thought experiment on two entangled particles which showed that quantum states
cannot in all situations be complete descriptions of physical reality. The EPR argument,
as subsequently modified by David Bohm (1951), goes as follows. Imagine the
singlet-spin state of two spin $\half$ particles

\begin{equation}
\ket{\Psi} = \frac{1}{\sqrt 2} \left ( \ket{\uparrow}\ket{\downarrow} -
\ket{\downarrow}\ket{\uparrow}\right ),
\label{singlet}
\end{equation}
where the single particle kets $\ket{\uparrow}$ and  $\ket{\downarrow}$ denote spin up
and spin down with respect to some chosen direction. This state is spherically
symmetric and the choice of the direction does not matter. The two particles, which we
label A and B, are emitted from a source and fly apart. After they are sufficiently
separated so that they do not interact with each other we can predict with certainity 
the x component of spin of particle A by measuring the x component of spin of particle B.
This is because the total spin of the two particles is zero and the spin components of
the two particles must have opposite values. The measurement performed on particle B
does not disturb particle A (by locality) therefore the x component of spin is an element
of reality according to the EPR criterion. By the same argument and by the spherical
symmetry of state $\ket{\Psi}$ the y, z, or any other spin components are also elements
of reality. However, since there is no quantum state of a spin $\half$ particle in which
all components of spin have definite values the quantum description of reality is not
complete. 

The EPR programme asked for a different description of quantum reality but until John
Bell's (1964) theorem it was not clear whether such a description was possible and if so
whether it would lead to different experimental predictions. Bell showed that the EPR 
propositions about locality, reality, and completeness are incompatibile with some 
quantum mechanical predictions involving entangled particles. The contradiction is
revealed by deriving from the EPR programme an experimentally testable inequality
which is violated by certain quantum mechanical predictions. Extension of Bell's original
theorem by John Clauser and Michael Horne (1974) made experimental tests of the EPR
programme feasible and quite a few of them have been performed. The experiments have
supported quantum mechanical predictions. Does this prove Bohr right? Not at all! The
refutation of the EPR programme does not give any credit to the Copenhagen
interpretation and simply shows that there is much more to `reality', `locality' and
`completeness'  than the EPR envisaged (for a contemporary realist's approach see, for
example, (Penrose 1989, 1994)).

What does it all have to do with data security? Surprisingly, a lot!  It turns out that the
very trick used by Bell to test the conceptual foundations of quantum theory can protect
data transmission from eavesdroppers! Perhaps it sounds less surprising when one
recalls again the EPR definition of an element of reality: ``If, without in any way
disturbing a system, we can predict with certainty the value of a physical quantity, then
there exists an element of physical reality corresponding to this physical quantity". If
this particular physical quantity is used to encode binary values of a cryptographic key
then all an eavesdropper wants is an element of reality corresponding to the encoding
observable (well, at least this was my way of thinking about it back in 1990).  Since
then several experiments have confirmed the `practical' aspect of Bell's theorem making
it quite clear that a border between blue sky and down-to-earth research is quite blurred.

\section{Quantum key distribution}

The quantum key distribution which I am going to discuss here is based on distribution
of entangled particles (Ekert 1991). Before I describe how the system works let me
mention that quantum cryptography does not have to be based on quantum entanglement.
In fact quite different approach based  on partial indistinguishibility of non-orthogonal
state vectors, pioneered by Stephen Wiesner (1983), and subsequently by Charles Bennett
and Gilles Brassard (1984), preceeded the entanglement-based quantum cryptography.
Entanglement, however, offers quite a broad repertoire of additional tricks such as, for
example, `quantum privacy amplification' (Deutsch et al. 1996) which makes the
entanglement-based key distribution secure and operable even in presence of
environmental noise.

The key distribution is performed via a quantum channel which consists of a source that
emits pairs of spin $\half$ particles in the singlet state as in Eq.(\ref{singlet}). The
particles fly apart along the z-axis towards the two legitimate users of the channel,
Alice and Bob, who, after the particles have separated, perform measurements and
register spin components along one of three directions, given by unit vectors
$\vec a_i$ and $\vec b_j\ \ (i,j=1,2,3)$, respectively for Alice and Bob. For simplicity
both $\vec a_i$ and $\vec b_j$ vectors lie in the x-y plane, perpendicular to the
trajectory of the particles, and are characterized by azimuthal angles: $\phi^a_1=0,
\phi^a_2={1\over 4}\pi,
\phi^a_3={1\over 2}\pi$ and $ \phi^b_1={1\over 4}\pi, \phi^b_2={1\over 2}\pi,
\phi^b_3={3\over 4}\pi$. Superscripts \lq\lq a" and \lq\lq b" refer to Alice's and Bob's
analysers respectively, and the angle is measured from the vertical x-axis. The users
choose the orientation of the analysers randomly and independently for each pair of the
incoming particles.  Each measurement, in  ${\half}\hbar$ units, can yield two results,
+1 (spin up) and -1 (spin down), and can potentially reveal one bit of information.

The quantity
\begin{equation} 
E(\vec a_i, \vec b_j) = P_{\scriptscriptstyle + +}(\vec a_i, \vec b_j) +
P_{\scriptscriptstyle - -}(\vec a_i,
\vec b_j) - P_{\scriptscriptstyle + -}(\vec a_i, \vec b_j) -
P_{\scriptscriptstyle - +}(\vec a_i, \vec b_j)
\end{equation}

is the correlation coefficient of the measurements performed by Alice along
$\vec a_i$ and by Bob along $\vec b_j$. Here $P_{\pm \pm} (\vec a_i, \vec b_j)$ denotes
the probability that result ${\pm} 1 $ has been obtained along $\vec a_i$ and ${\pm}1$
along $\vec b_j$.  According to the  quantum rules 
\begin{equation}
E(\vec a_i, \vec b_j) = - \vec a_i \cdot \vec b_j. 
\end{equation}

For the two pairs of analysers of the same orientation ($\vec a_2$, $\vec b_1$ and
$\vec a_3,\vec b_2$) quantum mechanics predicts total anticorrelation of the results
obtained by Alice and Bob: $E(\vec a_2,\vec b_1) = E(\vec a_3, \vec b_2) = -1$.

One can define quantity $S$ composed of the correlation coefficients for which Alice and
Bob used analysers of different orientation

\begin{equation} 
S=E(\vec a_1,\vec b_1)-E(\vec a_1,\vec b_3)+E(\vec a_3,\vec b_1)+E(\vec a_3,\vec
b_3). 
\label{defs}
\end{equation}
This is the same $S$ as in the generalised Bell theorem proposed by Clauser, Horne,
Shimony, and Holt (1969) and known as the CHSH inequality. Quantum mechanics requires

\begin{equation} 
S=-2{\sqrt 2}.
\label{qs}
\end{equation}
 
After the transmission has taken place, Alice and Bob can announce in public the
orientations of the analysers they have chosen for each particular measurement and
divide the measurements into two separate groups: a first group for which they used
different orientation of the analysers, and a second group for which they used the same
orientation of the analysers.  They discard all measurements in which either or both of
them failed to register a particle at all.  Subsequently Alice and Bob can reveal publicly
the results they obtained but within the first group of measurements only. This allows
them to establish the value of $S$, which if the particles were not directly or indirectly
`` disturbed" should reproduce the result of Eq.(\ref{qs}).  This assures the
legitimate users that the results they obtained within the second second group of
measurements are anticorrelated and can be converted into a secret string of bits ---
the key.

An eavesdropper, Eve, cannot elicit any information from the particles while in transit
from the source to the legitimate users, simply because there is no information encoded
there! The information  ``comes into being" only after the legitimate users perform
measurements and communicate in public afterwards. Eve may try to substitute her own
prepared data for Alice and Bob to misguide them, but as she does not know which
orientation of the analysers will be chosen for a given pair of particles there is no good
strategy to escape being detected. In this case her intervention will be equivalent to
introducing elements of {\it physical reality} to the spin components and will lower $S$
below its `quantum' value. Thus the Bell theorem can indeed expose eavesdroppers.

\section{Quantum eavesdropping}

The key distribution procedure described above is somewhat idealised.  The problem is
that there is in principle no way of distinguishing entanglement with an eavesdropper
(caused by her measurements) from entanglement with the environment caused by
innocent {\em noise}, some of which is presumably always present. This implies that all
existing protocols which do not address this problem are, strictly speaking, inoperable
in the presence of noise, since they require the transmission of messages to be
suspended whenever an eavesdropper (or, therefore, noise) is detected. Conversely, if we
want a protocol that is secure in the presence of noise, we must find one that allows
secure transmission to continue even in the presence of eavesdroppers. To this end, one
might consider modifying the existing protocols by reducing the statistical confidence
level at which Alice and Bob accept a batch of qubits. Instead of the extremely high
level envisaged in the idealised protocol, they would set the level so that they would
accept most batches that had encountered a given level of noise. They would then have to
assume that some of the information in the batch was known to an eavesdropper. It
seems reasonable that classical privacy amplification (Bennett 1995) could then be used
to distill, from large numbers of such qubits, a key in whose security one could have an
any desired level of confidence. However, no such scheme has yet been proved to
be secure. Existing proofs of the security of classical privacy amplification apply only
to classical communication channels and classical eavesdroppers. They do not cover the
new eavesdropping strategies that become possible in the quantum case: for instance,
causing a quantum ancilla to interact with the encrypted message, storing the ancilla
and later performing a measurement on it that is chosen according to the data that Alice
and Bob exchange publicly. The security criteria for this type of eavesdropping has only
recently been analysed (Gisin and Huttner 1996, Fuchs et al. 1997, Cirac and Gisin 1997).

The best way to analyse eavesdropping in the system is to adopt the scenario that is
most favourable for eavesdropping, namely where Eve herself is allowed to prepare all
the pairs that Alice and Bob will subsequently use to establish a key. This way we take
the most conservative view which attributes all disturbance in the channel to
eavesdropping even though most of it (if not all) may be due to an innocent
environmental noise. 

Let us start our analysis of eavesdropping in the spirit of the Bell theorem and
consider a simple case in which Eve knows precisely which particle is in which
state. Following (Ekert 1991) let us assume that Eve prepares each particle in the EPR
pairs separately so that each individual particle in the pair has a well defined spin in
some direction. These directions may vary from pair to pair so we can say that she 
prepares with probability $ p(\vec n_a,\vec n_b)$ Alice's particle in state $\ket{\vec
n_a}$ and Bob's particle in state
$\ket{\vec n_b}$, where
$\vec n_a$ and $\vec n_b$ are two unit vectors describing the spin orientations. This
kind of preparation gives Eve total control over the state of {\em individual} particles.
This is the case where Eve will always have the edge and Alice and Bob should abandon
establishing the key; they will learn  about it by estimating $|S|$ which in this case will
always be smaller than $\sqrt 2$. To see this let us write the density operator for each
pair as

\begin{equation}
\rho = \int p(\vec n_a, \vec n_b) \proj{\vec n_a} \otimes \proj{\vec n_b} \: d\vec n_a
d\vec n_b .
\label{mix}
\end{equation}

Equation (\ref{defs}) with appropriately modified correlation coefficients reads

\begin{eqnarray}
S =\int p(\vec n_a, \vec n_b) d\vec n_a d\vec n_b & &
        [(\vec a_1\cdot\vec n_a)(\vec b_1\cdot\vec n_b) -   
          (\vec a_1\cdot\vec n_a)(\vec b_3\cdot\vec n_b)+ \nonumber \\
	 & &(\vec a_3\cdot\vec n_a)(\vec b_1\cdot\vec n_b)+ 
          (\vec a_3\cdot\vec n_a)(\vec b_3\cdot\vec n_b)],
\end{eqnarray}

and leads to

\begin{equation}
S=\int p(\vec n_a, \vec n_b) d\vec n_a d\vec n_b [{\sqrt 2} \vec n_a \cdot \vec n_b]
\end{equation}
which implies
\begin{equation}
-{\sqrt 2} \leq S \leq {\sqrt 2},
\end{equation}
for any state preparation described by the probability distribution $p(\vec n_a, \vec
n_b)$.

Clearly Eve can give up her perfect control of quantum states of individual particles in
the pairs and entangle at least some of them. If she prepared all the pairs in perfectly
entangled singlet states she would loose all her control and knowledge about Alice's and
Bob's data who can then easily establish a secret key. This case is unrealistic because
in practice Alice and Bob will never register $|S|=2\sqrt 2$ . However, if Eve  prepares
only partially entangled pairs then it is still possible for Alice and Bob to establish the
key with an absolute security provided they use a {\it Quantum Privacy Amplification}
algorithm (QPA) (Deutsch et al. 1996).  The case of partially entangled pairs, $\sqrt 2\le
|S|\le 2\sqrt 2 $, is the most important one and in order to claim that we have an
operable key distribution scheme we have to prove that the key can be established in this
particular case. Skipping technical details I will present only the main idea behind the
QPA, details can be found in (Deutsch et al. 1996).

Firstly, note that any two particles that are jointly in a pure state cannot be entangled
with any third physical object.  Therefore any procedure that delivers EPR-pairs in pure
states must also have eliminated the entanglement between any of those pairs and any
other system. The QPA scheme is based on an iterative quantum algorithm which, if
performed with perfect accuracy, starting with a collection of EPR-pairs in mixed
states, would discard some of them and leave the remaining ones in states converging to
the pure singlet state. If (as must be the case realistically) the algorithm is performed
imperfectly, the density operator of the pairs remaining after each iteration will not
converge on the singlet but on a state close to it, however, the degree of entanglement
with any eavesdropper will nevertheless continue to fall, and can be brought to an
arbitrary low value. The QPA can be performed by Alice and Bob at distant locations by a
sequence of local unitary operations and measurements which are agreed upon by
communication over a public channel and could be implemented using technology that is
currently being developed (c.f. (Turchette et al. 1995)).

The essential element of the QPA procedure is the `entanglement purification' scheme,
the idea originally proposed by Charles Bennett, Gilles Brassard, Sandu Popescu,
Benjamin Schumacher, John Smolin, and Bill Wootters (1996). It has been shown recently
that any partially entangled states of two-state particles can be purified (Horodecki et
al 1997). Thus as long as the density operator cannot be written as a mixture of product
states, i.e. is not of the form (\ref{mix}), then  Alice and Bob may outsmart Eve!

Finally let me mention that quantum cryptography today is more than a theoretical
curiosity. Experimental work in Switzerland (Muller et al. 1995), in the U.K. (Townsend et
al. 1996), and in the U.S.A. (Hughes et al. 1996, Franson et al. 1995) shows that quantum
data security should be taken seriously!

\section{Public key cryptosystems}

In the late 1970s Whitfield Diffie and Martin Hellman (1976) proposed an interesting
solution to the key distribution problem. It involved two keys,
one public key $e$ for encryption and one private key $d$ for decryption:
\begin{equation}
\hat E_e(P) = C,  \; {\rm and\; }\; \hat D_d(C)=P.
\end{equation}
As I have already mentioned, in these systems users do not need to agree on any key
before they start sending  messages to each other. Every user has his own two keys; the
public key is publicly announced and the private key is kept secret. Several public-key
cryptosystems have been proposed since 1976; here we concentrate our attention on the
most popular one, which was already mentioned in Section 1, namely the RSA (Rivest et
al. 1979) . 

If Alice wants to send a secret message to Bob using the RSA system the first thing she
does is look up Bob's personal public key in a some sort of yellow pages or an RSA
public key directory. This consists of a pair of positive integers $(e,n)$. The integer $e$
may be relatively small, but $n$ will be gigantic, say a couple of hundred digits long.
Alice then writes her message as a sequence of numbers using, for example, our digital
alphabet from Section 1. This string of numbers is subsequently divided into blocks such
that each block when viewed as a number $P$ satisfies $P\le n$. Alice encrypts each $P$
as
\begin{equation}
\hat E(P) = C= P^e \bmod n.
\end{equation}
and sends the resulting cryptogram to Bob who can decrypt it by calculating
\begin{equation}
\hat D(C) = P= C^d \bmod n.
\end{equation}
Of course, for this system to work Bob has to follow a special procedure to generate both
his private and public key:
\begin{itemize}
\item He begins with choosing two large (100 or more digits long) prime numbers $p$ and
$q$, and a number $e$ which is relatively prime to both $p-1$ and $q-1$.
\item He then calculates $n=pq$ and finds $d$ such that $ed = 1\bmod (p-1)(q-1)$. This
equation can be easily solved, for example, using the extended Euclidean algorithm for
the greatest common divisor.~\footnote{Fortunately an easy and
very efficient algorithm to compute the greatest common divisor has been known since
300 BC. This truly `classical' algorithm is described in Euclid's {\it Elements}, the oldest
Greek treatise in mathematics to reach us in its entirety. Knuth (1981) provides an
extensive discussion of various versions of Euclid's algorithm.}
\item He releases to the public $n$ and $e$ and keeps $p$, $q$, and $d$ secret.
\end{itemize}

The mathematics behind the RSA is a lovely piece of number theory which goes back to
the XVI century when a French lawyer Pierre de Fermat discovered that if a prime $p$
and a positive integer $a$ are relatively prime, then
\begin{equation}
a^{p-1} =1 \bmod p.
\end{equation} 
A century later, Leonhard Euler found the more general relation
\begin{equation}
a^{\phi(n)} =1 \bmod n,
\end{equation} 
for relatively prime integers $a$ and $n$. Here $\phi (n)$ is Euler's $\phi$ function
which counts the number of positive integers smaller than $n$ and coprime to $n$.
Clearly for any prime integer such as $p$ or $q$ we have $\phi (p)= p-1$ and $\phi (q)=
q-1$; for $n=pq$ we obtain $\phi (n) = (p-1)(q-1)$. Thus the cryptogram $C=P^e \bmod n$
can indeed be decrypted by $C^d \bmod n = P^{ed} \bmod n$ because $ed = 1\bmod \phi
(n)$; hence for some integer $k$ 
\begin{equation}
P^{ed} \bmod n = P^{k\phi (n) +1} \bmod n = P.
\end{equation}

For example, let us suppose that Roger's public key is $(e,n) = (179,
571247)$.~\footnote{Needless to say, number $n$ in this example is too small to
guarantee security, do not try this public key with Roger.}   He generated it following the
prescription above choosing $p=773$, $q=739$ and $e=179$. The private key $d$ was
obtained by solving $179 d = 1\bmod 772\times 738$ using the extended Euclidean
algorithm which yields $d=515627$. Now if we want to send Roger encrypted ``HELLO
ROGER." we first use our digital alphabet from Section 1 to obtain the plaintext which
can be written as the following sequence of six digit numbers
\begin{equation}
021908 \quad\quad 071414 \quad\quad160708 \quad\quad 231503.
\end{equation}
Then we encipher each block $P_i$ by computing $C_i=P_i^e \bmod n$;  e.g. the first block
$P_1=021908$ will be eciphered as
\begin{equation}
P_1^e \bmod n = 021908^{179} \bmod 571247  =  540561 = C_1,
\end{equation}
and the whole message is enciphered as:
\begin{equation}
540561 \quad\quad 447313 \quad\quad 033313 \quad\quad 555657.
\end{equation}
The cryptogram $C$ composed of blocks $C_i$ can be send over to Roger. He can then
decrypt each block using his private key $d=515627$, e.g. the first block is decrypted as
\begin{equation}
540561^{515627} \bmod 571247  = 021908 = P_1.
\end{equation}

In order to recover plaintext $P$ from cryptogram $C$, an outsider, who knows $C, n,$
and  $e$, would have to solve the congruence 
\begin{equation}
P^e \bmod n = C, 
\end{equation}
for example, in our case,
\begin{equation}
P_1^{179} \bmod 571247 = 540561,
\end{equation}
which is hard, that is it is not known how to compute the solution efficiently when $n$
is a large integer (say $200$ decimal digits long or more). However, if we know the prime
decomposition of $n$ it is a piece of cake to figure out the private key $d$; we simply
follow the key generation procedure and solve the congruence $ed = 1\bmod (p-1)(q-1)$.
This can be done efficiently even when $p$ and $q$ are very large.  Thus, in principle,
anybody who knows $n$ can find $d$ by factoring $n$, but factoring big $n$ is a hard
problem. What does ``hard" mean ?

\section{Fast and slow algorithms}

Difficulty of factoring grows rapidly with the size, i.e. number of digits, of a number we
want to factor.  To see this take  a number $n$ with $l$ decimal digits ($n\approx 10^l$) 
and try to factor it by dividing it by $2,3,\ldots\sqrt n$ and checking the remainder. In
the worst case you may need approximately $\sqrt n = 10^{l/2}$ divisions to solve the
problem - an exponential increase as a function of $l$.  Now imagine a computer capable
of performing $10^{10}$ divisions per second.  The computer can then factor any number
$n$, using the trial division method, in about $\sqrt n/10^{10}$ seconds. Take a
$100\/$-digit number $n$, so that $n\approx 10^{100}$. The computer will factor this
number in about $10^{40}$ seconds, much longer than $10^{17}$ seconds - the estimated
age of the Universe! 

Skipping details of computational complexity I only mention that there is a rigorous way
of defining what makes an algorithm fast (and efficient) or slow (and impractical) (see,
for example (Welsh 1988)). For an algorithm to be considered fast, the time it takes to
execute the algorithm must increase no faster than a polynomial function of the size of
the input. Informally think about the input size as the total number of bits needed to
specify the input to the problem, for example, the number of bits needed to encode the
number we want to factorise. If the best algorithm we know for a particular problem has
execution time (viewed as a function of the size of the input) bounded by a polynomial
then we say that the problem belongs to class {\sf P}. Problems outside class {\sf P} are
known as hard problems. Thus we say, for example, that multiplication is in {\sf P}
whereas factorisation is apparently not in {\sf P} and that is why it is a hard problem.
We can also design non-deterministic algorithms which may sometimes produce
incorrect solutions but have the property that the probability of error can be made
arbitrarily small. For example, the algorithm may produce a candidate factor $p$ of the
input $n$ followed by a trial division to check whether $p$ really is a factor or not. If
the probability of error in this algorithm is $\epsilon$ and is independent of the size of
$n$ then by repeating the algorithm $k$ times, we get an algorithm which will be
successful with probability $1-\epsilon^k$ (i.e. having at least one success). This can be
made arbitrarily close to $1$ by choosing a fixed $k$ sufficiently large.

There is no known efficient classical algorithm for factoring even if we allow it to be
probabilistic in the above senses.  The fastest algorithms run in time roughly of order
$\exp ((\log n)^{1/3})$ and would need a couple of billions years to factor a 200-digit
number. It is not known whether a fast classical algorithm for factorisation exists or
not --- none has yet been found.

It seems that factoring big numbers will remain beyond the  capabilities of any realistic
computing devices and unless we come up with an efficient factoring algorithm the
public-key cryptosystems will remain secure.  Or will they? As it turns out we know
that this is not the case; the classical, purely mathematical, theory of computation is
not complete simply because it does not describe all physically possible computations.
In particular it does not describe computations which can be  performed by quantum
devices. Indeed, recent work in quantum computation shows that a quantum computer, at
least in principle, can efficiently factor large integers (Shor 1994).

\section{Quantum computers}

Quantum computers can compute faster because they can accept as the input not a single
number but a coherent superposition of many different numbers and subsequently
perform a computation (a sequence of unitary operations)  on all of these numbers
simultaneously. This can be viewed as a massive parallel computation, but instead of
having many processors working in parallel we have only one quantum processor
performing a computation  that affects all numbers in a superposition i.e. all components
of the input state vector.  

The exponential speed-up of quantum computers takes place at the very beginning of
their computation. Qubits, i.e. physical systems which can be prepared in one of the two
orthogonal states labelled as $\ket{0}$ and $\ket{1}$ or in a superposition of the two,
can store superpositions of many `classical' inputs. For example, the equally weighted
superposition of $\ket{0}$ and $\ket{1}$ can be prepared by taking a qubit initially in
state $\ket{0}$ and applying to it transformation {\bf H} (also known as the Hadamard
transform) which maps
\begin{eqnarray}
\ket{0} & \longrightarrow & \frac{1}{\sqrt 2} (\ket{0} + \ket{1}),\\
\ket{1} & \longrightarrow & \frac{1}{\sqrt 2} (\ket{0} - \ket{1}),
\label{tran}
\end{eqnarray}
If this transformation is applied to each qubit in a register composed of two qubits it
will generate the superposition of four numbers
\begin{eqnarray}
\ket{0}\ket{0}& \longrightarrow &\frac{1}{\sqrt 2}\left(\ket{0} + \ket{1}\right)
\frac{1}{\sqrt 2}\left(\ket{0} +
\ket{1}\right)\\
& & = \half (\ket{00} + \ket{01} + \ket{10} + \ket{11}).
\end{eqnarray}
where $00$ can be viewed as binary for $0$, $01$ binary for $1$, $10$ binary for $2$ and
finally $11$ as binary for $3$. In general a quantum register composed of $l$ qubits can
be prepared in a superposition of $2^l$ different numbers (inputs) with only $l$ 
elementary operations. This can be written, in decimal rather than in binary notation,  as
\begin{equation}
\ket{0}\longrightarrow 2^{-l/2} \sum_{x=0}^{2^l-1} \ket{x}.
\end{equation}
Thus $l$ elementary operations  generate exponentially many, that is $2^l$ different
inputs!

The next task is to process all the inputs in parallel within the superposition by a
sequence of unitary operations. Let us describe now how quantum computers compute
functions. For this we will need two quantum registers of length $l$ and $k$. Consider a
function
\begin{equation}
f:\, \{0,1,...\, 2^l-1\}\longrightarrow\{0,1,...\, 2^k-1\}.
\end{equation}

A classical computer computes $f$ by evolving each labelled input,
$0,1,...,\, 2^l-1$ into a respective labelled output, $f(0), f(1),...,\, f(2^l-1)$. Quantum
computers, due to the unitary (and therefore reversible) nature of their evolution,
compute functions in a slightly different way. In order to compute functions which are
not one-to-one and to preserve the reversibility of computation, quantum computers have
to keep the record of the input. Here is how it is done. The first register is loaded with
value $x$ i.e. it is prepared in state $\ket{x}$, the second register may initially contain
an arbitrary number $y$. The function evaluation is then determined by an appropriate
unitary evolution of the two registers,
\begin{equation}
\ket{x}\ket{y}\stackrel{U_f}{\longrightarrow}\ket{x}\ket{y + f(x)}.
\end{equation}
Here $y+f(x)$ means addition modulo the maximum number of configurations of the
second register, i.e. $2^k$ in our case.

The computation we are considering here is not only reversible but also quantum and we
can do much more than computing values of $f(x)$ one by one. We can prepare a
superposition of all input values as a single state and by running the computation $U_f$
{\em only once}, we can compute {\em all} of the $2^l$ values $f(0), \ldots ,f(2^l-1)$,
(here and in the following we ignore the normalisation constants),

\begin{equation}
\sum_{x=0}^{2^l-1} \ket{x}\ket{y}
\stackrel{U_f}{\longrightarrow} \ket{y+f}=\sum_{x=0}^{2^l-1} \ket{x}\ket{y+f(x)}.
\end{equation}

It looks too good to be true so where is the catch? How much information about $f$ does
the state
\begin{equation}
\ket{f} = \ket{0}\ket{f(0)} +\ket{1}\ket{f(1)}+\ldots
+\ket{2^l-1}\ket{f(2^l-1)}
\end{equation}
really contain?

Unfortunately no quantum measurement can extract all of the $2^l$ values
$f(0), f(1),\ldots ,f(2^l-1)$ from $\ket{f}$. If we measure the two registers after the
computation $U_f$ we register one output $\ket{x}\ket{y+f(x)}$ for some value $x$.
However, there are measurements that provide us with information about joint
properties of all the output values $f(x)$, such as, for example, periodicity, without
providing any information about particular values of $f(x)$. Let us illustrate this with
a simple example.

Consider a Boolean function $f$ which maps $\{0,1\}\rightarrow \{0,1\}$. There are
exactly four functions of this type: two constant functions ($f(0)=f(1)=0$ and
$f(0)=f(1)=1$) and two balanced functions ($f(0)=0, f(1)=1$ and $f(0)=1, f(1)=0$). Is it
possible to compute function $f$ {\em only once} and to find out whether it is constant or
balanced i.e. whether the binary numbers $f(0)$ and $f(1)$ are the same or different?
N.B. we are not asking for particular values $f(0)$ and $f(0)$ but for a global property of
$f$.

Classical intuition tells us that we have to evaluate both $f(0)$ and $f(1)$, that is to
compute $f$ twice. This is not so. Quantum mechanics allows us to perform the trick
with a single function evaluation.  We simply take two qubits, each qubit serves as a
single qubit register, prepare the first qubit in state $\ket{0}$ and the second in state
$\ket{1}$  and compute 
\begin{eqnarray}
\ket{0}\ket{1} &\longrightarrow & (\ket{0}+\ket{1})(\ket{0}-\ket{1}) \longrightarrow
\nonumber\\
& \longrightarrow & \ket{0} (\ket{f(0)} - \ket{1+f(0)}) + \ket{1}(\ket{f(1)} -
\ket{1+f(1)}).
\end{eqnarray}
We start with transformation {\bf H} applied both to the first and the second qubit, 
followed by the function evaluation. Here $1+f(0)$ denotes addition modulo 2 and simply
means taking the negation of $f(0)$.  At this stage, depending on values $f(0)$ and
$f(1)$, we have one of the four possible states of the two qubits.  We apply {\bf H} again
to the first and the second qubit and evolve the four states as follows

\begin{eqnarray}
(\ket{0} +\ket{1})(\ket{0} - \ket{1}) & \longrightarrow &  + \ket{0} \ket{1},\\
(\ket{0} +\ket{1})(\ket{1} - \ket{0}) & \longrightarrow &  - \ket{0} \ket{1},\\
\ket{0}(\ket{0} - \ket{1}) +\ket{1}(\ket{1} - \ket{0}) & \longrightarrow &  +\ket{1}
\ket{1},\\
\ket{0}(\ket{1} - \ket{0}) +\ket{1}(\ket{0} - \ket{1}) & \longrightarrow &  - \ket{1}
\ket{1}.
\end{eqnarray}
The second qubit returns to its initial state $\ket{1}$ but the first qubit contains the
relevant information. We measure its bit value --- if we register `0' the function is
constant if we register `1' the function is balanced !

This example, due to Richard Cleve, Artur Ekert and Chiara Macchiavello (1996), is an improved
version of the first quantum algorithm proposed by David Deutsch (1985) and communicated to
the Royal Society by Roger Penrose. (The original Deutsch algorithm provides the correct
answer with probability 50\% .) Deutsch's paper laid the foundation for the new field of
quantum computation. Since then quantum algorithms have been steadily improved and in 1994
Peter Shor came up with the efficient quantum factoring algorithm which, at least in theory,
leads us directly to quantum cryptanalysis.

\section{Quantum code-breaking}

Shor's quantum factoring of an integer $n$ is based on calculating the period of the
function $F_n(x) = a^x\bmod n$ for a randomly selected integer $a$ between $0$ and
$n$. It turns out that for increasing powers of $a$, the remainders form a repeating
sequence with a period which we denote $r$. Once $r$ is known the factors of
$n$ are obtained by calculating the greatest common divisor of $n$ and $a^{r/2}\pm 1$.

Suppose we want to factor $15$ using this method.  Let $a=11$. For increasing $x$ the
function $11^x\bmod 15$ forms a repeating sequence $1,11,1,11,1,11,\ldots$. The period
is $r=2$, and $a^{r/2}\bmod 15 =11$. Then we take the greatest common divisor of $10$
and
$15$, and of $12$ and $15$ which gives us respectively $5$ and $3$, the two factors of
$15$. Classically calculating $r$ is at least as difficult as trying to factor $n$; the
execution time of calculations grows exponentially with number of digits in $n$.
Quantum computers can find $r$ in time which grows only as a cubic function of the
number of digits in $n$. 

To estimate the period $r$ we prepare two quantum registers; the first register, with $l$
qubits, in the equally weighted superposition of all numbers it can contain, and the
second register in state zero.  Then we perform an arithmetical operation that takes
advantage of quantum parallelism by computing the function $F_n(x)$ for each number
$x$ in the superposition. The values of $F_n(x)$ are placed in the second register so that
after the computation the two registers become entangled:
\beq
\sum_x\ket{x}\ket{0} \longrightarrow \sum_x \ket{x}\ket{F_n(x)}
\eeq
Now we perform a measurement on the second register. We measure each qubit and
obtain either "0" or "1" . This measurement yields value  $F_n(k)$ (in binary notation) for
some randomly selected $k$. The state of the first register right after the measurement,
due to the periodicity of $F_n(x)$, is a coherent superposition of all states $\ket{x}$
such that $x=k, k+r, k+2r,\ldots$, i.e. all
$x$ for which $F_n(x)=F_n(k)$. The periodicity in the probability amplitudes in the first
register cannot be simply measured because the offset i.e. the value
$k$ is randomly selected by the measurement. However the state of the first register
can be subsequently transformed  via a unitary operation which effectivelly removes the
offset and modifies the period in the probability amplitudes from $r$ to a multiple of
$2^l/r$. This operation is known as the quantum Fourier transform (QFT) and can be
written as

\begin{equation}
\mbox{\rm QFT}_s : \ket{x} \longmapsto 2^{-l/2} \sum_{y=0}^{2^l-1} \exp(2\pi
iac/2^l)\: \ket{y}.
\label{qftdef}
\end{equation}

After QFT the first register is ready for the final measurement which yields with high
probability an integer which is the best whole approximation of a multiple of $2^l/r$ i.e.
$x = k  2^l/r$ for some integer $k$.  We know the measured value $x$ and the size of the
register $l$ hence if $k$  and $r$ are coprime we can determine $r$ by canceling $x/2^l$
down to an irreducible fraction and taking its denominator. Since the probability that $k$
and $r$ are coprime is sufficiently large (greater than $1/\log r$ for large $r$) this
gives an efficient randomized algorithm for determination of $r$. More detailed
description of Shor's algorithm can be found in (Shor 1994) and in (Ekert and Jozsa 1996).

Let me mention in passing that there exists a direct quantum attack on RSA which does
not require factoring, but employs Shor's algorithm to determine the order of cryptogram
modulo $n$ (Mosca and Ekert 1997).

An open question has been whether it would ever be practical to build physical devices to
perform such computations, or whether they would forever remain theoretical
curiosities. Quantum computers require a coherent, controlled evolution for a period of
time which is necessary to complete the computation. Many view this requirement as an
insurmountable experimental problem, however, the technological progress  may prove
them wrong (see review papers e.g. (DiVincenzo 1995; Ekert and Jozsa 1996;
Lloyd 1993, 1995)).  When the first quantum factoring devices are built the security of
classical public-key cryptosystems will vanish. But, as was pointed by Roger Penrose,
by the time we acquire desk-top quantum computers we will probably be able to
construct quantum public-key cryptosystems with security based on quantum rather than
classical computational complexity. Meanwhile thumbs up for quantum cryptography.

\section{Concluding remarks}

In the last decade quantum entanglement became a sought-after physical resource which
allows us to perform qualitatively new types of data processing. Here I have described
the role of entanglement in connection with data security and have skipped
many other fascinating application such as quantum teleportation (Bennett et al. 1993),
quantum dense coding (Bennett and Wiesner 1992), entanglement swapping (Zukowski et
al. 1993), quantum error correction (Shor 1995; Steane 1996; Ekert and Macchiavello
1996; Calderbank and Shor 1996; Bennett et al. 1996, Laflamme et al. 1996), fault tolerant 
quantum computing
(Shor 1996; DiVincenzo and Shor 1996) and only mentioned in passing the entanglement
purification (Bennett et al.  1996)  and quantum privacy amplification (Deutsch et al
1996). There is much more to say about some peculiar features of two- and
many-particle entanglement and there is even an interesting geometry behind it such as,
for example, the `magic dedocahedra' (Penrose 1994).  Even the `simplest' three-particle
entangled states, the celebrated GHZ states of three qubits (Greenberger et al. 1989)
such as
$\ket{000}\pm \ket{111}$, have interesting properties; the three particles are all
together entangled but none of the two qubits in the triplet are entangled. That is, the
pure GHZ state of three particles is entangled as a whole but the reduced density
operator of any of the pairs is separable. This is very reminicent of some geometric
constructions such as ``Odin's triangle" or ``Borromean rings" (Aravind 1997).
\begin{figure}[!ht]
\vspace*{0.2cm}
\center{\psfig{height=6cm,file=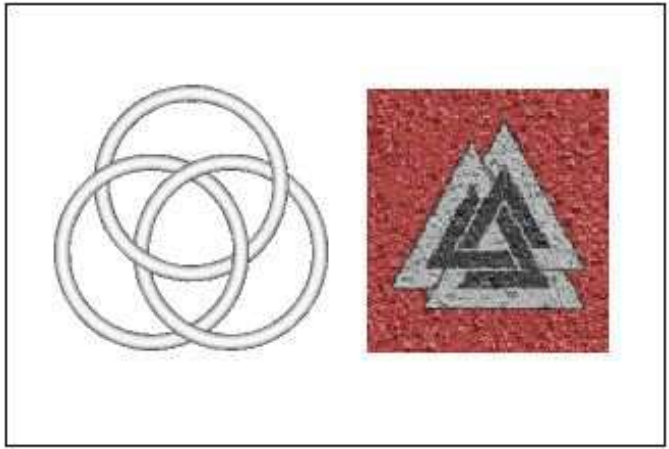}}
\vspace*{5mm}
\caption{The Borromean rings consist of three `entangled' rings with the property
that if any one of them is removed, then all three separate. The name Borromean comes
from the Italian Borromeo family who used them for their coat of arms. Odin's triangle 
was found in picture-stones on Gotland, an island in the Baltic sea off the southeast
coast of Sweden. These are dated around the ninth century and are thought to tell tales
from the Norse myths. To the Norse people of Scandinavia the interlocked triangles are
known as ``Odin's triangle" of the ``Walknot" (or ``valknut" -- the knot of the slain). The
symbol was also carved on the bedposts used in their burials at sea
(see, for example, http://sue.csc.uvic.ca/~cos/venn/borromean.html).}
\label{odin}
\end{figure}
This and many other interesting properties of multi-particle entanglement have been
recently studied by Sandu Popescu (1997) and (in a disguised form as quantum error
correction) by Rob Calderbank, Eric Rains, Peter Shor, and Neil Sloane (1996). Despite a
remarkable progress in this field it seems to me that we still know very little about the
nature of quantum entanglement; we can hardly agree on how to classify, quantify and
measure it (but c.f. (Horodecki 1997; Vedral et al.1997). Clearly we should be prepared
for even more surprises both in understanding and in utilisation of this precious quantum
resource.

In this quest to understand the quantum theory better and better Roger Penrose has
played a very prominent role in defending the realist's view of the
quantum world. I would like to thank him for this.

\section{Acknowledgements}
I am greatly indebted to David DiVincenzo, Chiara Macchiavello and Michele Mosca for
their comments and help in preparation of this manuscript. The author is supported by the
Royal Society, London. This work was supported in part by the European TMR Research
Network ERP-4061PL95-1412,  Hewlett-Packard and Elsag-Bailey.


\begin{thebibliography}{99}

\bibitem{Ara} Aravind, P.K. (1997) {\em Borromean entanglement of the GHZ state} in {\em
Potentiality, entanglement and passion-at-a-distance}, edited by R.Cohen, M.Horne and
J.Stachel, Kluwer Academic Publishers. 

\bibitem{BDEJ}  Barenco, A., Deutsch, D., Ekert, A. and Jozsa, R. (1995) {\em
Phys.~Rev.~Lett.} {\bf 74},~4083.

\bibitem{Bell} Bell, J.S. (1964) {\em Physics }{\bf 1}, 195.

\bibitem{phystoday} Bennett, C.H. (1996) {\em Phys. Today} {\bf 48}, 27.

\bibitem{BB84} Bennett, C.~H. and Brassard, G. (1984) in  \lq\lq Proc. IEEE Int. Conference
on Computers, Systems and Signal Processing", IEEE, New York, (1984).

\bibitem{BW} Bennett, C.H. and Wiesner, S.J. (1992) {\em Phys. Rev. Lett.} {\bf 69} 2881. 

\bibitem{BBCM} Bennett, C.H., Brassard, G., Cr\'epeau, C. and Maurer, U.M. (1995) {\em
IEEE Trans. Inf. Th.} {\bf IT-41}, 1915.

\bibitem{teleport} Bennett, C.~H., Brassard, G., Cr\'epeau, C., Jozsa, R., Peres, A. and
Wootters, W.K. (1993) {\em Phys. Rev. Lett.} {\bf 70}, 1895.

\bibitem{pur} Bennett, C.~H., Brassard, G., Popescu, S., Schumacher, B.,
  Smolin, J. and  Wootters, W.~K. (1996) {\em Phys.~Rev.~Lett.} {\bf 76}, 722.

\bibitem{BDSW} Bennett, C.H., DiVincenzo, D.P., Smolin, J.A., and Wootters, W.K. (1996)
Phys. Rev. A {\bf 54}, 3824. 

\bibitem{Bohm} Bohm, D. (1951) {\em Quantum Theory}, Prentice-Hall.

\bibitem{cald} Calderbank, A.R. and Shor, P.W. (1996) {\em Phys. Rev. A} {\bf 54}, 1098.

\bibitem{cald2} Calderbank, A.R., Rains, E.M., Shor, P.W. and Sloane, N.J.A. (1966) {\em
Quantum Error Correction via Codes over GF(4)} quant-ph/9608006. 

\bibitem{CG} Cirac, J.I. and Gisin N. (1997) {\em Coherent eavesdropping strategies for
the 4 state quantum cryptography protocol} quant-ph/9702002.

\bibitem{CH} Clauser, J.F. and Horne, M.A.  (1974) {\em Phys. Rev. D} {\bf 10}, 526.

\bibitem{CHSH} Clauser, J.F. and Horne, M.A., Shimony, A. and Holt, R.A. (1969) {\em Phys.
Rev. Lett.} {\bf 23}, 880.

\bibitem{CEMM} Cleve, R., Ekert, A. and Macchiavello, C. (1996) - during long afternoon
discussions on quantum algoritms and Californian wine at the Santa Barbara Workshop on Quantum
Computation and Decoherence. More general analysis of quantum algorithms will be provided in
Cleve, R., Ekert, A., Macchiavello, C. and Mosca, M. {\em Quantum Algorithms Revisited} (in
preparation). 

\bibitem{D85} Deutsch, D. (1985) {\em Proc. R. Soc. London~A} {\bf 400}, 97.

\bibitem{QPA} Deutsch, D., Ekert,A., Jozsa, R., Macchiavello, C., Popescu, S. and Sanpera,
A. (1996) {\em Phys. Rev. Lett.} {\bf 77} 2818.

\bibitem{PubK} Diffie, W. and Hellman, M.E. (1976) {\em IEEE Trans.~Inf.~Theory} {\bf
IT-22}, 644.

\bibitem{DiVi1}  DiVincenzo, D.P. (1995) {\em Science} {\bf 270}, 255.

\bibitem{DiVi2} DiVincenzo, D.P. and Shor, P.W. (1996) {\em Phys. Rev. Lett.} {\bf 77},
3260.

\bibitem{EPR} Einstein, A., Podolsky, B. and Rosen N. (1935) {\em Phys. Rev.} {\bf 47},
777.

\bibitem{Ekert} Ekert,  A. (1991) {\em Phys.~Rev.~Lett.} {\bf 67}, 661. 

\bibitem{EJ} Ekert, A. and Jozsa, R. (1996) {\em Rev. Mod. Phys.} {\bf 68}, 733.

\bibitem{EM} Ekert, A. and Macchiavello, C. (1996) {\em Phys. Rev. Lett.} {\bf 77}, 2585.

\bibitem{FGGNP} Fuchs, C.A., Gisin, N., Griffiths, R.B., Niu, C-S. and Peres A. (1997) {\em
Optimal eavesdropping in quantum cryptography I} quant-ph/9701039.

\bibitem{Franson} Franson, J.D. and Jacobs, B.C. (1995) {\em Electron. Lett.} {\bf 31}, 232.

\bibitem{GB} Gisin, N. and Huttner, B.  (1996) {\em Quantum cloning, eavesdropping, and
Bell's inequality}, quant-ph/9611041.


\bibitem{GHZ}  Greenberger, D.~M., Horne, M. and Zeilinger, A. 1989 Going beyond Bell's
theorem, in {\em Bell's Theorem, Quantum Theory, and Conceptions of the Universe}, ed.
by M.~Kafatos, Kluwer, Dordrecht pp.69-72.

\bibitem{hor} Horodecki, M., Horodecki, P. and Horodecki, R. (1997) {\em Phys. Rev. Lett.}
{\bf 78} 574.

\bibitem{Hugh} Hughes, R.J., Luther, G.G., Morgan, G.L., Peterson, C.G. and Simmons, C.
(1996) {\em Quantum cryptography over underground optical fibres}, Advances in
Cryptology - Proceedings of Crypto'96, Springer-Verlag. 

\bibitem{Kahn} Kahn, D. {\it The Codebreakers: The Story of Secret Writing\/},
Macmillan, New York (1967).

\bibitem{Knuth} Knuth, D.E. (1981) {\em The Art of Computer Programming, Volume 2/Seminumerical 
Algorithms} Addison-Wesley.

\bibitem{lafl} Laflamme, R., Miquel, C., Paz, J.P., and Zurek, W.H. (1996) {\em Phys. Rev. 
Lett.} {\bf 77}, 198.

\bibitem{lloyd1} Lloyd, S. (1993) {\em Science} {\bf 261}, 1569.

\bibitem{lloyd4} Lloyd, S. (1995), {\em Scient. Am.} {\bf 273}, 44.

\bibitem{Menezes} Menezes, A.J., van Oorschot, P.C. and Vanstone S.A. (1996) {\em
Handbook of Applied Cryptography} CRC Press. 

\bibitem{RSAdirect} Mosca, M. and Ekert, A. (1997) {\em A note on quantum attack on
RSA}, unpublished, available from the authors.

\bibitem{geneva} Muller, A. , Zbinden, H. and Gisin, N. (1995) {\it Nature}, {\bf 378}, 449.

\bibitem{Penrose1} Penrose, R. (1989) {\em The Emperor's New Mind}, Oxford University
Press.

\bibitem{Penrose2} Penrose, R. (1994) {\em Shadows of the Mind}, Oxford University
Press.

\bibitem{Popescu} Popescu, S. (1997) private communication.

\bibitem{RSA} Rivest, R., Shamir, A. and Adleman, L.{\it On Digital  Signatures and
Public-Key Cryptosystems}, MIT Laboratory for Computer  Science, Technical Report,
MIT/LCS/TR-212 (January 1979).

\bibitem{Schneier}Schneier, B. (1994) {\it Applied cryptography: protocols, algorithms,
and source code in C.\/}, John Wiley \& Sons.

\bibitem{Schrodinger} Schr\"odinger,  E. (1935) ``Die gegenw\"artige Situation in der 
Quantenmechanik", Naturwissenschaften, {\bf 23}, 807-812; 823-828; 844-849. English 
translation, ``The Present Situation in Quantum Mechanics", Proc. of  the American
Philosophical Society, {\bf 124}, 323-338 (1980); reprinted  in {\em Quantum Theory and
Measurement} edited by J.A.~Wheeler and  W.H.~Zurek, (Princeton, 1983) pp.152-167.

\bibitem{Shor} Shor, P.W.  (1994) in {\em Proceedings of the 35th Annual Symposium on
the Foundations of Computer Science}, edited by S.~Goldwasser (IEEE Computer Society
Press, Los Alamitos, CA), p. 124; Expanded version of this paper is available at LANL
quant-ph archive.

\bibitem{Shor1} Shor, P.W. (1995) {\em Phys. Rev. A}, {\bf 52}, 2493.

\bibitem{Shor2} Shor, P.W. (1996) {\em Fault-tolerant quantum computation}
quant-ph/9605011.

\bibitem{Steane} Steane, A. M. (1996) {\em Phys. Rev. Lett.} {\bf 77}, 793; Steane, A.M. 
{\em Proc. R. Soc. London A} {\bf 452}, 2551.

\bibitem{BT} Townsend, P.D., Marand, C., Phoenix, S.J.D., Blow, K.J. and Barnett, S.M.
(1996) {\em Phil. Trans. Roy. Soc. London A} {\bf 354}, 805.

\bibitem{turch}Turchette,  Q.A., Hood, C.J., Lange, ~W., Mabuchi, H. and Kimble, H.J. (1995)
{\em Phys.~Rev.~Lett. }{\bf 75}, 4710.

\bibitem{ved} Vedral, V., Plenio, M.P., Rippin, M.A., and Knight, P.L. (1997) {\em
Quantifying Entanglement} quant-ph/9702027 and {\em Phys. Rev. Lett.} (to appear).

\bibitem{Welsh} Welsh, D. {\it Codes and Cryptography\/}, Clarendon Press, Oxford
(1988)

\bibitem{Wiesner} Wiesner, S. (1983) {\em SIGACT News}, {\bf 15}, 78 (1983); 
original manuscript written {\em circa\/}~1970. 

\bibitem{swap} Zukowski, M.,  Zeilinger, A., Horne, M. and A.K.Ekert (1993) {\em Phys.
Rev. Lett.} {\bf 71}, 4287.

\end{thebibliography}
\end{document}